# Light bullets in moiré lattices

## YAROSLAV V. KARTASHOV[1,*]

[1]Institute of Spectroscopy, Russian Academy of Sciences, 108840, Troitsk, Moscow, Russia
*Corresponding author: Yaroslav.Kartashov@icfo.eu

We predict that photonic moiré lattices produced by two mutually twisted periodic sublattices in the medium with Kerr nonlinearity can support stable three-dimensional light bullets localized in both space and time. Stability of light bullets and their properties are tightly connected with the properties of linear spatial eigenmodes of moiré lattice that undergo localization-delocalization transition (LDT) upon increase of the depth of one of the sublattices forming moiré lattice, but only for twist angles corresponding to incommensurate, aperiodic moiré structures. Above LDT threshold such incommensurate moiré lattices support stable light bullets without energy threshold. In contrast, commensurate, or periodic, moiré lattices arising at Pythagorean twist angles, whose eigenmodes are delocalized Bloch waves, can support stable light bullets only above certain energy threshold. Moiré lattices below LDT threshold cannot support stable light bullets for our parameters. Our results illustrate that periodicity/aperiodicity of the underlying lattice is a crucial factor determining stability properties of the nonlinear three-dimensional states.

Spatiotemporal solitons or light bullets are unique self-sustained three-dimensional (3D) wavepackets that may form under conditions when anomalous group-velocity dispersion and diffraction are simultaneously exactly balanced by a focusing nonlinearity of the material [1]. Light bullets may exist as solutions of corresponding evolution equations in materials with different types of nonlinear response and in different geometries, but truly challenging issue from the experimental point of view is stability of such states. In particular, 3D light bullets are unstable [2,3] and may exhibit collapse in materials with ubiquitous cubic Kerr nonlinearity. Different approaches to realization of stable 3D light bullets were proposed [4-8]. They include utilization of quadratic, competing, and nonlocal nonlinearities, fast longitudinal modulations of parameters of the material, dissipative and higher-order effects, and, most notably, utilization of optical lattices. Thus, periodic waveguide arrays that support rich families of 2D spatial solitons [9-15], have been suggested also as a collapse-suppressing tool in 3D allowing to obtain stable solitons even in cubic optical medium [16-18]. Periodic arrays have been used for first experimental realization of the fundamental and vortex light bullets [19,20]. Various 3D states, including vortex bullets [21,22], have been reported in radially-periodic lattice structures [23] and graded-index fibers [24-26].

Stable solitons may form not only in periodic, but also in aperiodic lattices, where they were studied so far only in 2D settings [27-30]. A remarkable new class of optical lattices allowing continuous transformation between periodic and aperiodic geometries is represented by moiré lattices produced by two mutually twisted periodic sublattices [31]. It has been shown in [32,33] that LDT phenomenon [31,34] for 2D linear modes in such lattices also dramatically affects the formation of spatial solitons in them – qualitative transformation of the eigenmodes of the moiré lattice in the vicinity of Pythagorean twist angles, where it becomes periodic, was shown to lead to sharp variations of the threshold power for 2D soliton formation.

The goal of this Letter is to show that moiré lattices can support stable 3D light bullets and that localization properties of linear spatial eigenmodes of moiré structures directly impact energy threshold required for the formation of stable 3D light bullets. In particular, we show that incommensurate aperiodic moiré lattices support thresholdless light bullets that broaden in time, but remain spatially-localized in low-amplitude limit and that stability domains for bullets appear in our system only above LDT threshold.

We consider propagation of light pulses with central wavelength corresponding to the range where group velocity dispersion (GVD) is anomalous in focusing cubic medium with transverse shallow refractive index modulation in the form of moiré lattice. Propagation dynamics is governed by the nonlinear Schrödinger equation for the dimensionless light field amplitude $\psi$:

$$i\frac{\partial \psi}{\partial z} = -\frac{1}{2}\left(\frac{\partial^2 \psi}{\partial x^2} + \frac{\partial^2 \psi}{\partial y^2} + \frac{\partial^2 \psi}{\partial \tau^2}\right) - |\psi|^2\,\psi - \mathcal{R}(\mathbf{r})\psi, \qquad (1)$$

where $x, y$ are the scaled transverse coordinates, $\mathbf{r} = (x, y)$, $z$ is the propagation distance, $\tau = t - z/v_{gr}$ is the reduced time, $v_{gr}$ is the group velocity of the carrier wave, $\mathcal{R}(\mathbf{r}) = |p_1\mathcal{V}(\mathbf{S}\mathbf{r}) + p_2\mathcal{V}(\mathbf{r})|^2$ is the function describing moiré lattice created by superposition of two mutually rotated by angle $\theta$ [ $\mathbf{S}(\theta)$ is the operator of 2D rotation in the $(x, y)$ plane] square sublattices $\mathcal{V}(\mathbf{r}) = \cos(\Omega x) + \cos(\Omega y)$ with depths $p_{1,2}$ and $\Omega = 2$. For Pythagorean angles $\theta = \arctan[2mn/(m^2 - n^2)]$, $m, n \in \mathbb{N}$, associated with the Pythagorean triple $(m^2 - n^2, 2mn, m^2 + n^2)$ moiré lattice is exactly periodic, or commensurate [in the top row of Fig. 1(a) we show the lattice for $(m, n) = (2,1)$ but for the angle

$\theta = \arctan[(m^2 - n^2)/2mn]$ producing identical, but simply rotated lattice]. For all other angles moiré lattice is aperiodic, or incommensurate [see bottom row of Fig. 1(a) for example of aperiodic lattice]. Moiré lattices can be imprinted all-optically in photosensitive materials or inscribed with fs-laser in transparent dielectrics [8,9,13]. It has been shown in [31] that properties of spatial 2D modes supported by moiré lattices may qualitatively change upon increase of the depth of one of the sublattices forming the structure. To illustrate this, we omit nonlinearity and GVD in Eq. (1), calculate linear spatial eigenmodes of the lattice in the form $\psi = w(x,y)e^{ibz}$, where $b$ is the propagation constant, and plot in Fig. 1(b) the form-factor $\chi = (U^{-2}\iint |\psi|^4 d^2\mathbf{r})^{1/2}$, where $U = \iint |\psi|^2 d^2\mathbf{r}$, of the most lo-calized eigenmode with largest $b$ value as a function of the twist angle $\theta$ and sublattice depth $p_2$ (hereafter we fix $p_1 = 0.3$). One can see that for non-Pythagorean angles $\theta$ the modes of aperiodic moiré lattice exhibit LDT and become localized (their form-factor, inversely proportional to the width, acquires values $\sim 1$) for $p_2 > p_2^{cr} \approx 0.51$ [examples of modes below and above LDT are shown in the top and bottom rows of Fig. 1(c), respectively]. In contrast, for Pythagorean angles $\theta$, some of which are highlighted by the dashed lines in Fig. 1(b), the eigenmodes remain delocalized for any $p_2$, resulting in dips in $\chi(\theta, p_2)$ dependence. The dependence $\chi(\theta, p_2)$ is symmetric with respect to the angle $\theta = \pi/2$.

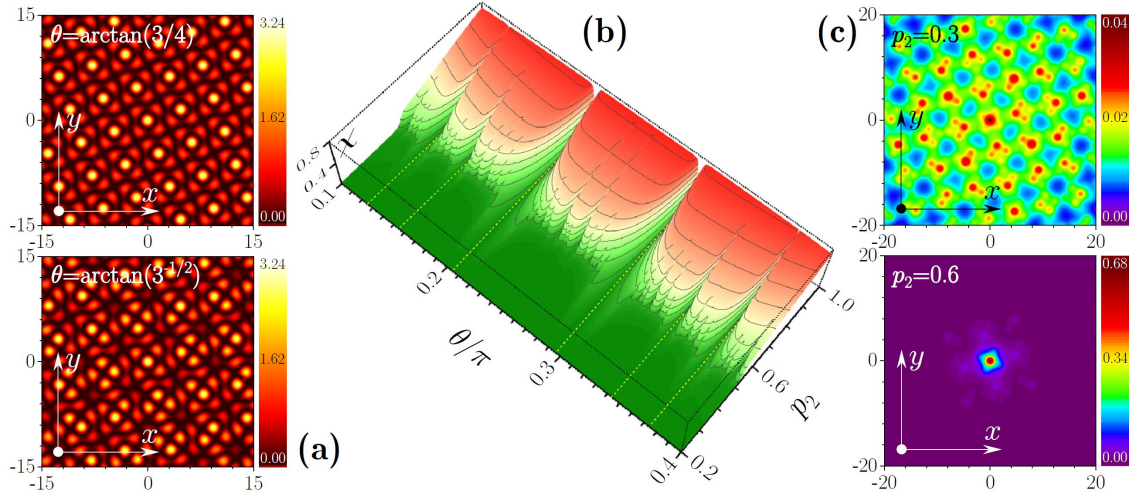

Fig. 1. (a) Color-fill plots showing commensurate (top) and incommensurate (bottom) moiré lattices $\mathcal{R}(\mathbf{r})$ with $p_1 = 0.3$, $p_2 = 0.6$. (b) Form-factor of the eigenmode of moiré lattice with largest propagation constant versus $\theta$ and $p_2$ at $p_1 = 0.3$. Yellow dashed lines indicate some of the Pythagorean angles corresponding to commensurate moiré lattices. (c) Eigenmodes with the largest propagation constant supported by moiré lattice with $p_1 = 0.3$, $\theta = \arctan(3^{-1/2})$ for different $p_2$ values.

Such qualitative change of the spectrum of spatial 2D modes of the system finds its manifestation in qualitative change of the behavior of 3D light bullet solutions of Eq. (1), at least close to cutoffs for their existence that are determined by eigenvalues of corresponding spatial modes. To illustrate this we calculate exact 3D nonlinear states in the form $\psi = w(x, y, \tau)e^{ibz}$ and plot in Fig. 2(a) the dependencies of their energy $E = \iiint |\psi|^2 d^2\mathbf{r} d\tau$ on propagation constant $b$ for various $p_2$ values and for two angles: Pythagorean $\theta = \arctan(3/4)$ (red curves) and non-Pythagorean $\theta = \arctan(3^{-1/2})$ (black curves) ones. Light bullets exist above the cutoff value $b = b_{co}$ defined by the eigenvalue of spatial mode of moiré lattice from which bifurcation occurs. In the cutoff, where peak amplitude of the bullet $\max|\psi|$ vanishes, the duration $W_\tau = 2(E^{-1}\iiint \tau^2 |\psi|^2 d^2\mathbf{r} d\tau)^{1/2}$ of the light bullet always dramatically increases [Fig. 2(c)]. In contrast, its spatial width $W_x = 2(E^{-1}\iiint x^2 |\psi|^2 d^2\mathbf{r} d\tau)^{1/2} = W_y$ diverges only when light bullet bifurcates from delocalized in space states, but remains finite if bifurcation occurs from localized modes above LDT threshold. This substantially affects corresponding $E(b)$ dependencies. Thus, for $p_2 < p_2^{cr}$, below LDT threshold, when lattice does not support localized modes, the energy monotonically grows with decrease of $b$ irrespectively of the value of twist angle $\theta$. For $p_2 > p_2^{cr}$, i.e. above LDT threshold, for non-Pythagorean angles there exists the interval of propagation constants, where energy decreases with decrease of

$b$ and vanishes at $b \to b_{co}$ (in this case the bullet remains spatially localized). In contrast, for the same $p_2$ value, but for Pythagorean angle, the energy first decreases, but then grows very close to the cutoff resulting in nonmonotonic $E(b)$ dependence because spatial profile of the bullet approaches delocalized Bloch wave of periodic lattice. Zooms of $E(b)$ curves in the vicinity of cutoff showing qualitatively different behavior of light bullets in periodic and aperiodic moiré lattices are shown in Fig. 2(b),(d). These dependencies imply that aperiodic moiré lattices enable excitation of stable (at the segments with positive slope $dE/db > 0$, as per Vakhitov-Kolokolov criterion, applicable in our case) light bullets even at $E \to 0$, in contrast to periodic lattices, where the above condition can be met only in finite interval of energies and always above certain minimal energy. The difference between $E(b)$ dependencies in periodic and aperiodic lattices is most pronounced for $p_2$ values near LDT threshold [curves 3 and 4 in Fig. 2(a)]. With increase of $p_2$ a local minimum in the $E(b)$ curve for Pythagorean angle is acquired at smaller and smaller energies, so that curves for different angles become closer (curves 5).

The examples of profiles of the light bullets close to the cutoff are shown in Fig. 2(e) and 2(f) at $p_2 > p_2^{cr}$. For Pythagorean angle light bullet expands drastically in space acquiring multiple peaks, while for non-Pythagorean angle it remains spatially relatively well-local-ized (depending on how far is $p_2$ from LDT threshold $p_2^{cr}$). With increase of $b$ the bullet contracts in space and time [Fig. 2(g),(h)] until it shrinks, in the high-amplitude limit, to practically single maximum of the lattice. In this regime, in our continuous model the energy of the bullet starts decreasing with increase of $b$. This limits the interval of propagation constants, where the slope $dE/db$ is positive and where one can expect stability of the light bullets. We found similar behavior of light bullet families for other angles –above LDT threshold a local minimum is present in the $E(b)$ curve for Pythagorean angles [at least for $(m, n) = (3, 2)$ and $(m, n) = (4, 3)$ lattices], while for non-Pythagorean ones the energy vanishes in the cutoff indicating on the existence of thresholdless states. We studied stability of bullets by propagating them directly with small-scale noise added into input field distributions. Since our bullets are the simplest fundamental self-sustained states, the Vakhitov-Kolokolov criterion is applicable for them: it predicts stability of states at the segments with positive slope $dE/db$ and instability of the states at the segments with $dE/db < 0$.

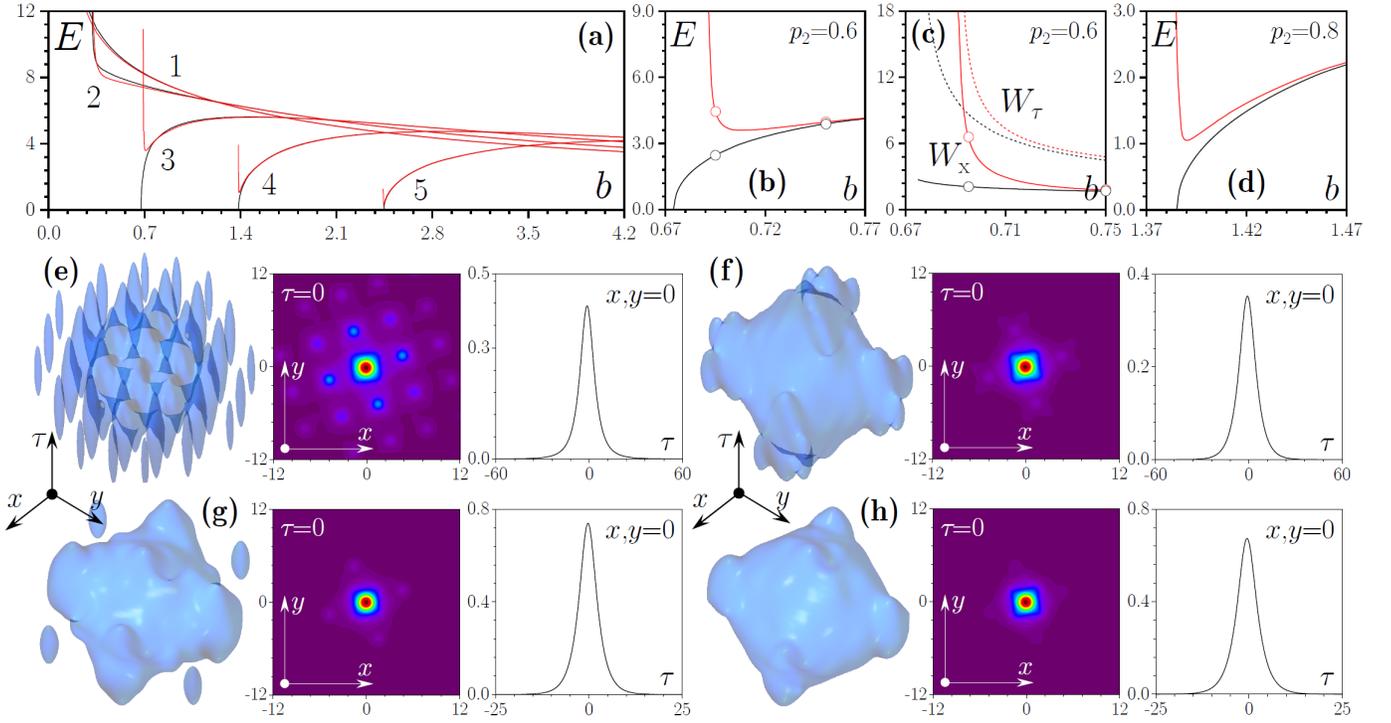

Fig. 2. (a) Energy of light bullet versus $b$ for $p_2 = 0.2$ (curves 1), $0.4$ (2), $0.6$ (3), $0.8$ (4), and $1.0$ (5). Black curves correspond to $\theta = \arctan(3^{-1/2})$, red curves to $\theta = \arctan(3/4)$. (b), (d) Zoom of $E(b)$ dependencies around cutoffs. (c) Spatial (solid lines) and temporal (dashed lines) widths of bullet versus $b$ at $p_2 = 0.6$. Isosurface plots of $|\psi|$ at the level $0.005$, spatial $|\psi|$ distributions at $\tau = 0$, and temporal profiles at $x, y = 0$ for light bullets with (e) $b = 0.695$, $\theta = \arctan(3/4)$, (f) $b = 0.695$, $\theta = \arctan(3^{-1/2})$, (g) $b = 0.750$, $\theta = \arctan(3/4)$, and (h) $b = 0.750$, $\theta = \arctan(3^{-1/2})$, corresponding to dots in panel (b).

Figures 3(a) and 3(b) illustrate unstable propagation of the bullet in periodic and stable evolution of the bullet in aperiodic moiré lattice for the same near-cutoff propagation constant values $b = 0.695$, respectively. These states correspond to the left red and black dots in Fig. 2(b). One can see that due to instability development, the state from $dE/db < 0$ segment in periodic lattice [Fig. 3(a)] exhibits large oscillations, accompanied by contraction of the central high-amplitude part of the bullet. In contrast, stable light bullet in aperiodic lattice shows minimal amplitude oscillations in the course of propagation [Fig. 3(b)]. Another example of stable propagation of the light bullet with $b = 1.15$, approximately from the middle of stability segment $b \in [0.674, 1.476]$ in aperiodic lattice is illustrated in Fig. 3(c). This bullet is already more localized in space and time. Its counterpart with the same $b$ in periodic moiré lattice is stable too. Finally, in Fig. 3(d) we show dynamics of the unstable high-amplitude strongly localized state with $b = 2$ outside the above stability domain. For selected noise realization the amplitude of the bullet rapidly drops at the initial stage and starts oscillating around new, lower value. This is accompanied by considerable spatiotemporal reshaping and emission of additional pulses.

Summarizing, we have reported on the existence of stable light bullets in moiré lattices, both in periodic and aperiodic configurations. We have shown that localization properties of corresponding spatial eigenmodes are crucial for the behavior and stability of nonlinear 3D states, especially for lattice depths near the LDT threshold, where bullet properties substantially differ in periodic and aperiodic structures.

**Funding:** Russian Science Foundation (grant 21-12-00096).

**Disclosures:** The author declares no conflicts of interest.

**Acknowledgements:** Fruitful discussions with Dr. S. K. Ivanov are gratefully acknowledged.

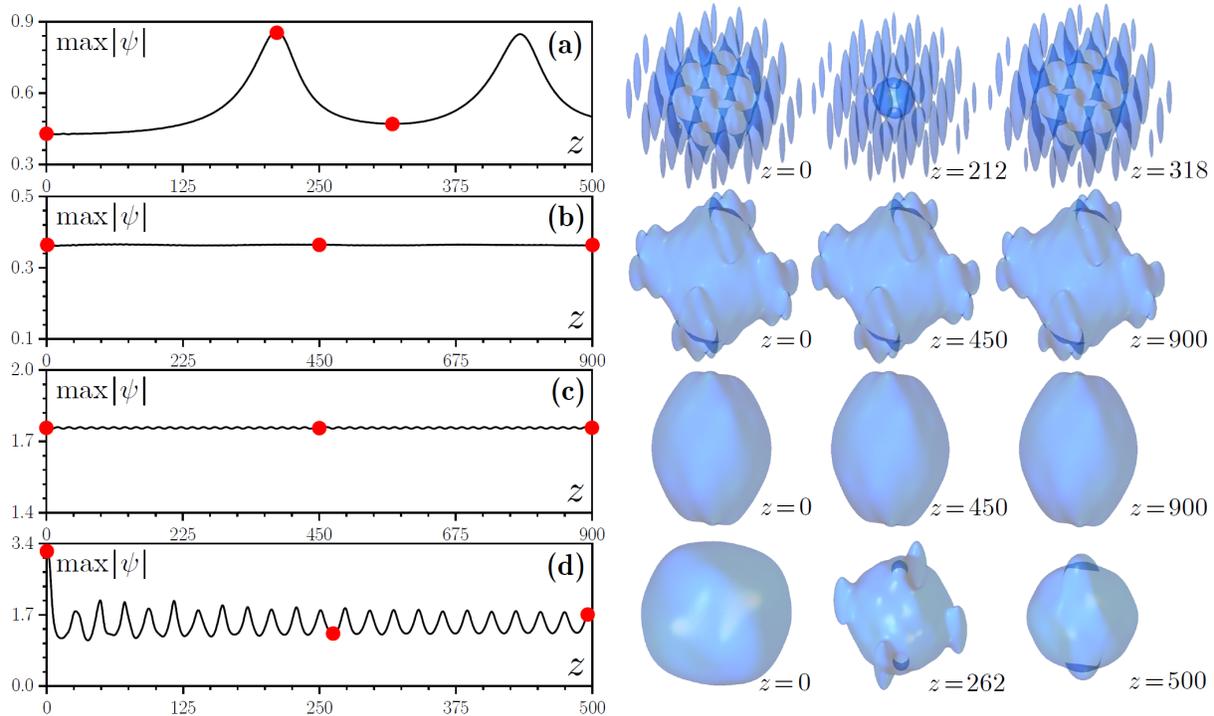

Fig. 3. (a) Decay of the unstable light bullet with $b = 0.695$ in the lattice with $\theta = \arctan(3/4)$, (b) stable propagation at $b = 0.695$, $\theta = \arctan(3^{-1/2})$, (c) stable propagation at $b = 1.15$, $\theta = \arctan(3^{-1/2})$, and (d) decay at $b = 2$, $\theta = \arctan(3^{-1/2})$. Left column shows dependence of peak amplitude of the light bullet on distance, right column shows isosurfaces at $|\psi| = 0.005$ level for distances corresponding to the red dots in the left column. In all cases $p_1 = 0.3$, $p_2 = 0.6$.

# References with titles